%% file: main.tex
\documentclass[conference,a4paper]{IEEEtran}

\renewcommand\IEEEkeywordsname{Index Terms}

\usepackage{amsmath,amssymb,bm}
\usepackage{graphicx}
\usepackage{booktabs}
\usepackage[table]{xcolor}
\usepackage{arydshln}
\usepackage{cite}
\usepackage[hyphens]{url}
\usepackage[bookmarks=false]{hyperref}
\usepackage[nolist]{acronym}
\usepackage{balance}
\usepackage{flafter}
\input{corporateColours}

\usepackage{tikz}
\usetikzlibrary{arrows.meta,positioning,calc,fit,shapes.geometric,
                decorations.pathreplacing,spy,backgrounds}
\usepackage{pgfplots}
\pgfplotsset{compat=1.18}
\usepgfplotslibrary{groupplots}

\input{macros}

\newcommand{\Ccal}{\mathcal{C}}
\newcommand{\Lch}{\boldsymbol{\ell}_{\mathrm{ch}}}

\newcommand{\tauv}{\vecs{\tau}}
\newcommand{\phiv}{\vecs{\phi}}
\newcommand{\Bcal}{\mathcal{B}}
\newcommand{\Lcal}{\mathcal{L}}

\IEEEoverridecommandlockouts
\pgfdeclarelayer{edgelayer}
\pgfsetlayers{background,edgelayer,main}

\makeatletter
\tikzset{
    database top segment style/.style={draw},
    database middle segment style/.style={draw},
    database bottom segment style/.style={draw},
    database/.style={
        path picture={
            \path [database bottom segment style]
                (-\db@r,-0.5*\db@sh) 
                -- ++(0,-1*\db@sh) 
                arc [start angle=180, end angle=360,
                    x radius=\db@r, y radius=\db@ar*\db@r]
                -- ++(0,1*\db@sh)
                arc [start angle=360, end angle=180,
                    x radius=\db@r, y radius=\db@ar*\db@r];
            \path [database middle segment style]
                (-\db@r,0.5*\db@sh) 
                -- ++(0,-1*\db@sh) 
                arc [start angle=180, end angle=360,
                    x radius=\db@r, y radius=\db@ar*\db@r]
                -- ++(0,1*\db@sh)
                arc [start angle=360, end angle=180,
                    x radius=\db@r, y radius=\db@ar*\db@r];
            \path [database top segment style]
                (-\db@r,1.5*\db@sh) 
                -- ++(0,-1*\db@sh) 
                arc [start angle=180, end angle=360,
                    x radius=\db@r, y radius=\db@ar*\db@r]
                -- ++(0,1*\db@sh)
                arc [start angle=360, end angle=180,
                    x radius=\db@r, y radius=\db@ar*\db@r];
            \path [database top segment style]
                (0, 1.5*\db@sh) circle [x radius=\db@r, y radius=\db@ar*\db@r];
        },
        minimum width=2*\db@r + \pgflinewidth,
        minimum height=3*\db@sh + 2*\db@ar*\db@r + \pgflinewidth,
    },
    database segment height/.store in=\db@sh,
    database radius/.store in=\db@r,
    database aspect ratio/.store in=\db@ar,
    database segment height=0.1cm,
    database radius=0.25cm,
    database aspect ratio=0.35,
    database top segment/.style={
        database top segment style/.append style={#1}},
    database middle segment/.style={
        database middle segment style/.append style={#1}},
    database bottom segment/.style={
        database bottom segment style/.append style={#1}}
}
\makeatother

\begin{document}
\setlength{\columnsep}{0.21in}

\title{{When to Stop? Dynamic Early Termination of Sequential Ensembles}}

\author{\IEEEauthorblockN{Paul Bezner\textsuperscript{\dag}, Felix
Krieg\textsuperscript{\dag}, and Stephan ten Brink}
\IEEEauthorblockA{Institute of Telecommunications, University of Stuttgart, Germany\\
\{bezner, krieg, tenbrink\}@inue.uni-stuttgart.de}
\thanks{\textsuperscript{\dag}P. Bezner and F. Krieg contributed equally to this work.}
\thanks{This work is supported by the German Federal Ministry of Research, Technology and Space (BMFTR) within the project Open6GHub+ (grant no. 16KIS2406)}
\thanks{LLM agents from Anthropic and OpenAI were used to write source code for
this work and to assist with language editing.}}

\maketitle
\suppressfloats[t]

\begin{acronym}
\acro{AED}{automorphism ensemble decoding}
\acro{AWGN}{additive white Gaussian noise}
\acro{BP}{belief propagation}
\acro{BPSK}{binary phase-shift keying}
\acro{CN}{check node}
\acro{CSI}{channel state information}
\acro{DAE}{dynamic automorphism ensemble}
\acro{DET}{dynamic ensemble termination}
\acro{FER}{frame error rate}
\acro{GE}{Gaussian elimination}
\acro{LDPC}{low-density parity-check}
\acro{LLR}{log-likelihood ratio}
\acro{MBBP}{multiple-bases belief propagation}
\acro{ML}{maximum likelihood}
\acro{MRB}{most-reliable basis}
\acro{NMSA}{normalized min-sum algorithm}
\acro{OSD}{ordered-statistics decoding}
\acro{PCM}{parity-check matrix}
\acro{QC}{quasi-cyclic}
\acro{RBE}{row-boosted ensemble}
\acro{SNR}{signal-to-noise ratio}
\acro{VN}{variable node}
\acro{SC}{successive-cancellation}
\end{acronym}

\begin{abstract}
\Ac{OSD} post-processing substantially improves the performance of \ac{BP}
ensembles, but it also removes their natural syndrome-based stopping
criterion.
The high computational complexity of ensemble decoding and \ac{OSD} can be reduced by sequential activation of ensemble members paired with an early termination.
Instead of a syndrome-based termination we therefore apply \ac{DET} to sequential ensembles: members are evaluated one at a time and decoding stops when the estimated risk that an unseen member would correct the current decision falls below an offline-fitted threshold.
With hardware-oriented design in mind, we instantiate the framework with layered normalized min-sum component decoders, a row-boosted ensemble, conditional low-order \ac{OSD} on the dual code, and a conventional low-cost decoder gate.
A decision tree estimates the residual risk from statistics of the current candidate list.
Member-dependent thresholds then target a prescribed \ac{FER} loss.
We validate this framework on five IEEE~802.11 \ac{LDPC} codes with
blocklengths \(648,1296\), and \(1944\) at rates \(1/2\) and \(5/6\).
At each code's design point, where the full \(64\)-member list reaches an
\ac{FER} of \(10^{-3}\), the \ac{DET} decoder processes \(1.04\) to
\(1.15\) members on average.
On the \((648,540)\) code, it reduces \ac{BP} work by \(39\times\) and
\ac{OSD} activations by \(18\times\) against the full list while retaining an
\ac{SNR} gain of about \(0.4\,\mathrm{dB}\) 

\end{abstract}

\acresetall
\section{Introduction}
\label{sec:introduction}
\Ac{LDPC} codes are commonly decoded by iterative message passing on a sparse
\ac{PCM}~\cite{tanner}. Layered scheduling reduces the
number of iterations by applying each updated check message
immediately~\cite{hocevarlayered}. However, a single trajectory can still
stagnate in a suboptimal fixed point. Extending that trajectory does not
resolve the issue, as a longer run creates no decoder diversity -- which could move the decoder out of a trapping set.
\begin{figure}[t]
    \centering
    \input{fig/model1}
    \vspace{-1.5em}
    \caption{{Risk-controlled sequential ensemble
  decoder used in this work. The layered \acs{BP} gate
  returns a syndrome-valid estimate immediately. On failure, the afterburner
  evaluates row-boosted members one at a time and conditionally refines their outputs by \mbox{\acs{OSD}-1}. The decision whether to stop or to process the next member is based on the residual-risk estimate.}}
 \vspace{-0.3cm}
    \label{fig:intro_fig}
\end{figure}
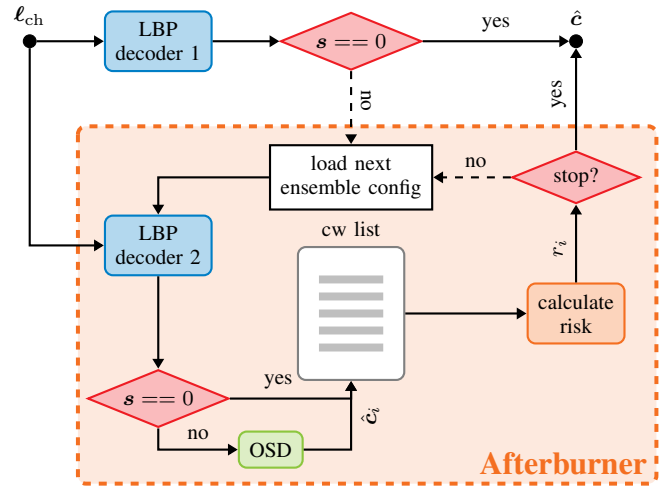
Ensemble decoders address this limitation by creating several decoding
trajectories. \Ac{MBBP} uses multiple parity-check
bases~\cite{hehn2010multiple} and \ac{AED} exploits code 
automorphisms~\cite{Chen_Cyclic_LDPC_AED,geiselhart2022breaking}. 
Reliability-based post-processing such as \ac{OSD} can further turn an invalid
\ac{BP} output into a valid
candidate~\cite{fossorier1995osd}. This improves the \ac{FER} performance of the decoder, however, it also changes the stopping problem, since a member with nonzero syndrome may still produce the best final candidate after \ac{OSD}-refinement.
Consequently, stopping at the first syndrome-valid output is not sufficient.
We instead apply \ac{DET} by estimating the probability of premature termination of the ensemble and fit thresholds for termination based on a prescribed relative \ac{FER} loss.
Each skipped member avoids both \ac{BP} iterations and a potential
\ac{OSD} activation.
On the \((1944,1620)\) code,
spending a \(64\)-fold larger iteration budget on a single trajectory improves
the \ac{FER} by roughly a factor of two, whereas distributing the same
worst-case budget over a post-processed \(64\)-member ensemble gains almost an order of magnitude. However, a direct implementation either instantiates many decoders in parallel, which is costly in area, or evaluates the
full list sequentially, which is costly in latency.
Sequential \ac{AED} with early stopping was recently proposed to solve this issue for \ac{SC} decoding of polar
codes~\cite{pillet2026dae}. This idea does not depend on the ensemble construction: process the members in a fixed order and continue only while the channel metric of the best current candidate exceeds an optimized threshold. 
We apply this early stopping principle to the \ac{RBE} introduced in~\cite{bezner2026rbe} and replace the single path metric by a risk of the ensemble being terminated prematurely. Note that the approach of \cite{pillet2026dae} is applicable to any ensemble and stopping criterion and hence, we refer to it as \ac{DET}:  plain \ac{DET}
thresholds the channel metric as in \cite{pillet2026dae}, risk-\ac{DET} calculates a score trying to match the imposed risk of terminating early.
Design choices such as the component decoder, diversity construction, and post-processor in this work are practical implementation choices~\cite{bezner2026rbe,zhang2025unfoldedOSD}.
For \ac{RBE} all ensemble members share one \ac{PCM} and only require a single amplified \ac{CN} respectively. Hence, sequential evaluation reuses a single decoder instance. The sequential implementation translates into saved area and energy rather than into parallel hardware.
To avoid paying the cost for the ensemble and the \ac{OSD} post-processing on frames that a single decoder already resolves, we place most of the ensemble
as an afterburner behind a conventional decoder
gate~\cite{wehn2016afterburner}. 
We fit the risk model once per code at a single optimization \ac{SNR}. To sustain a high throughput despite the variable per-frame runtime, the
decoder is combined with buffering and multicore
scheduling.
Hence, the
average decoder latency, which is reduced by sequential implementation, determines the achievable throughput.

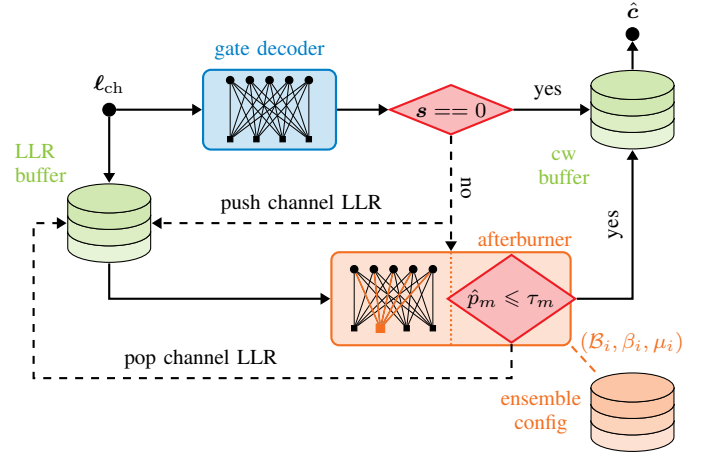
\begin{figure}[!t]
    \centering
    \input{fig/model2}
    \vspace{-2em}
    \caption{Two-core mapping of the decoder. The gate occupies one core,
    the sequential afterburner the other, and the \acs{LLR} buffer holds frames
    waiting for it. All members reuse \(\Hm\) and differ only in the boosted
    row and its gain; stopping parameters are fitted offline. Dashed arrows indicate control logic.}
    \label{fig:architecture}
\end{figure}

The central contribution of this work is a stopping framework for
sequential ensembles whose members always output valid codewords.
\begin{itemize}
    \item We formulate risk-controlled \ac{DET} in which premature termination is measured by the probability that an unseen member would correct the current decision (Sec.~\ref{sec:decoder}),
    \item provide a residual-risk score and use an offline, per-member threshold fit to a prescribed
    relative \ac{FER} loss, %
    \item construct a hardware-oriented gated sequential architecture that conditionally invokes post-processing, and maps the variable
    workload onto two buffered cores (Sec.~\ref{sec:decoder}),
    \item validate experimentally on IEEE~802.11 codes of varying block length and rate using a
    layered \ac{BP}, \ac{RBE}-64, and \mbox{\ac{OSD}-1} instantiation
    (Sec.~\ref{sec:results}).
\end{itemize}

\section{Preliminaries}
\label{sec:prelim}

\subsection{Code Family and Notation}

Throughout this work, vectors and matrices are denoted by boldface lowercase
(\(\cv\)) and uppercase (\(\Hm\)) letters, respectively. Estimates are marked by a hat, e.g., \(\hat\cv\).

Let \(\Ccal(n,k)\) be a binary \ac{LDPC} code of length \(n\), dimension
\(k\), and rate \(R=k/n\), defined by a \ac{PCM}
\(\Hm\in\mathbb{F}_2^{(n-k)\times n}\) as
\(\Ccal=\{\cv\in\mathbb{F}_2^n:\Hm\cv^{\mathsf T}=\zerov\}\). We denote the transmitted codeword by \(\cv\) 
and the corresponding decoder estimate by \(\hat\cv\). The
IEEE~802.11 codes are \ac{QC}. A base matrix of \(24\) block columns is lifted
by cyclic permutation matrices of size \(Z\in\{27,54,81\}\), which yields
\(n\in\{648,1296,1944\}\)~\cite{ieee80211}. 
We consider \(R\in\{1/2,5/6\}\) in this work.
\subsection{Channel Model}
The codeword is mapped to \ac{BPSK} symbols \(s_j=1-2c_j\). Over the
memoryless \ac{AWGN} channel, received samples are
\begin{equation}
    y_j = s_j + n_j,\qquad n_j\sim\mathcal{N}(0,\sigma^2),
    \label{eq:awgn}
\end{equation}
with noise variance \(\sigma^2=(2R\,E_\mathrm{b}/N_0)^{-1}\). The
corresponding channel \acp{LLR} \(\ell_{\mathrm{ch},j}=2y_j/\sigma^2\) are
collected in \(\Lch\) and initialize the decoders.

\section{Decoder Architecture}

\label{sec:decoder}

\subsection{Layered Normalized Min-Sum}

The constituent decoder is the layered \acf{NMSA}. Let \(N(i)\) denote
the set of \acp{VN} connected to \ac{CN} \(i\), let \(q_{j\rightarrow i}\)
denote the message from \ac{VN} \(j\) to \ac{CN} \(i\), and let
\(r_{i\rightarrow j}\) denote the message in the opposite direction. Each update starts from the extrinsic 
\begin{equation}
    q_{j\rightarrow i}=Q_j-r_{i\rightarrow j},
    \label{eq:extrinsic}
\end{equation}
where \(Q_j\) denotes the posterior \ac{LLR} of \ac{VN} \(j\), initialized as
\(Q_j=\ell_{\mathrm{ch},j}\). The check node update is
\begin{equation}
 r_{i\rightarrow j} =
 \alpha
 \prod_{j'\in N(i)\setminus j}\!\operatorname{sgn}
       (q_{j'\rightarrow i})
 \min_{j'\in N(i)\setminus j}|q_{j'\rightarrow i}|
 \label{eq:nmsa}
\end{equation}
with the normalization factor \(\alpha\), and the posterior update
\begin{equation}
        Q_j = q_{j\rightarrow i}+r_{i\rightarrow j}.
    \label{eq:posterior}
\end{equation}
A layer comprises the \(Z\) rows of one lifted check row. Hence, one iteration
is a sweep over all \((n-k)/Z\) layers, and at most \(I_{\max}\) iterations
are performed. After each iteration, we compute the syndrome
\(\sv=\Hm\hat\cv^{\mathsf T}\) of the hard decision \(\hat\cv\) and stop as
soon as \(\sv=\zerov\).

\subsection{Row-Boosted Members}

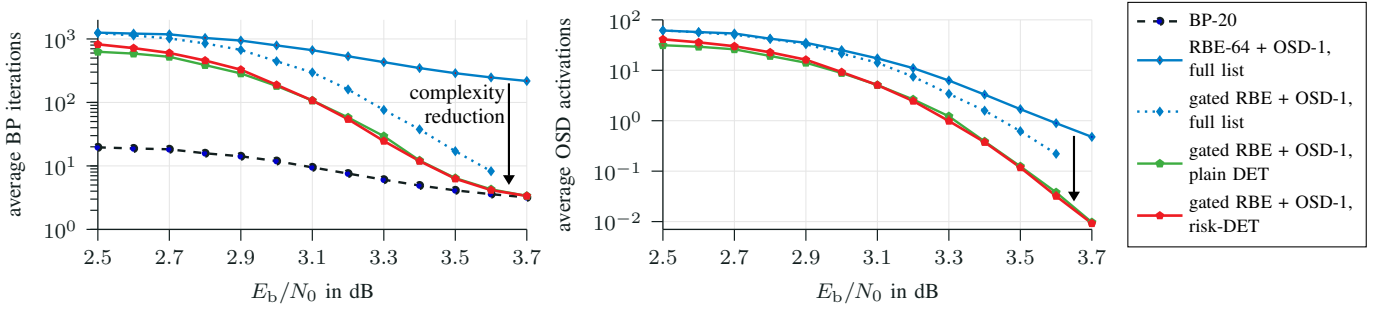
\begin{figure*}[!t]
    \centering
    \input{fig/wifi1944_layered_complexity.tikz}
    \vspace{-0.7cm}
    \caption{Average \acs{BP} iterations (left) and \mbox{\acs{OSD}-1}
    activations (right) per transmitted codeword against \ac{SNR} for the \((1944,1620)\)
    code. Plain \acs{DET} \textcolor{neuesgruen}{(green)} and risk-\acs{DET} \textcolor{rot}{(red)} are fitted at \(3.5\,\mathrm{dB}\) on the
    same dataset and with the same excess \ac{FER} budget.}
    \label{fig:complexity}
\end{figure*}
We use \ac{RBE} as diversity mechanism.
All \(M\) ensemble members operate on the same graph and use the same check
node update. Let \(m\in\{0,\dots,M-1\}\) index the members. Member \(m=0\) is
the plain decoder, which also serves as the gate, and is therefore included in
every average member count reported in Sec.~\ref{sec:results}. All other members are defined by a set \(\Bcal_m\) of boosted \acp{CN} and a gain \(\beta_m\). In this work, we use \(|\Bcal_m|=1\) and \(\beta_m=\beta=4\) \(\forall\, m\).
The check messages remain unscaled in memory, and only the posterior
increment of a boosted row is scaled.
Therefore, \eqref{eq:posterior} is replaced by
\begin{equation}
\begin{aligned}
 Q_j &\leftarrow Q_j+
 \gamma_i^{(m)}\bigl(r_{i\rightarrow j}^{\mathrm{new}}
                    -r_{i\rightarrow j}^{\mathrm{old}}\bigr),\\
 \gamma_i^{(m)}&=
 \begin{cases}
 \beta_m,&i\in\Bcal_m,\\
 1,&\text{otherwise},
 \end{cases}
\end{aligned}
 \label{eq:rowboost}
\end{equation}
where \(r^{\mathrm{new}}_{i\rightarrow j}\) and
\(r^{\mathrm{old}}_{i\rightarrow j}\) denote the check-to-variable message
after and before the update in \eqref{eq:nmsa}. Note that
\eqref{eq:rowboost} reduces to \eqref{eq:posterior} for
\(\gamma_i^{(m)}=1\), i.e., for every non-boosted row.

All candidates are validated against \(\Hm\). Hence, a member perturbs the
message trajectory without changing the code constraints, and no second
\ac{PCM} must be stored. For \(M=64\), the \(63\) singleton sets \(\Bcal_m\)
select \(63\) distinct rows among the minimum-degree rows of \(\Hm\). The
decoder therefore stores \(63\) row indices and a single gain. 

\subsection{Conditional OSD and List Selection}

\begin{figure}[!t]
    \centering
    \input{fig/wifi648_r56_fer.tex}
    \vspace{-0.7cm}
    \caption{
    FER performance for \((648,540)\) IEEE 802.11 LDPC code.
    At a FER of \(10^{-3}\) dynamic ensemble termination with truncated OSD-\(1\) provides a
    \(0.4\,\mathrm{dB}\) gain over BP followed by OSD-\(1\).
    }
    \label{fig:wifi648_r56_fer}
\end{figure}
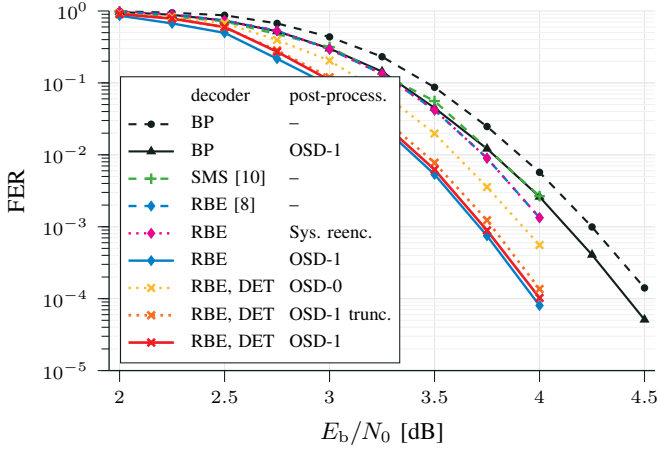

We use \ac{OSD} for post-processing. Applying \ac{OSD} to an ensemble decoder yields a larger gain than applying it
to a single decoder. However, the computational overhead scales with the ensemble size. With early termination we activate the next member, including \ac{OSD}, only as long as a further member is going to improve the decision to keep the complexity as low as possible.
If a member terminates with zero syndrome, its hard decision enters the
candidate list directly; otherwise, we apply \mbox{\ac{OSD}-1}  to its
posterior \acp{LLR}, sort the positions by reliability, obtain the \ac{MRB}
by \ac{GE}, and re-encode the order-zero word together with the \(k\)
weight-one test patterns~\cite{fossorier1995osd}.
An activation therefore costs one \ac{GE} and \(k+1\) re-encodings, whereas a member with a
valid output requires no activation at all. Truncation of test patterns can reduce complexity with minor \ac{FER} loss as shown in Fig.~\ref{fig:wifi648_r56_fer} where only $99<k$ test patterns are used. Performance for \mbox{\ac{OSD}-0} post-processing with no test patterns is also visualized as a lower-complexity alternative.
Note that \ac{OSD} can be implemented as reliability-based
syndrome decoding on \(\Hm\), the generator of the dual code, instead of
reducing the larger primal generator \(\Gm\)~\cite{fossorier1998dualOSD}.
The equivalent candidate is retained while \ac{GE} operates on only
\((n-k)\) rows.
Efficient unfolded-\ac{GE} architectures may further reduce
its implementation latency~\cite{zhang2025unfoldedOSD}.
Let \(\bv\) denote the channel hard decision, i.e., \(b_j=1\) if
\(L_{\mathrm{ch},j}<0\). Among all valid candidates
\(\tilde\cv\) collected in the list, the decoder retains the one that
minimizes the channel metric
\begin{equation}
 C(\tilde\cv) = \!\!\sum_{j:\,\tilde c_j\ne b_j}\!\! |L_{\mathrm{ch},j}| ,
 \label{eq:metric}
\end{equation}
that is, the total reliability of the positions in which the candidate
contradicts the channel. For a
memoryless binary-input channel with exact \acp{LLR}, minimizing
\eqref{eq:metric} is equivalent to \ac{ML} decoding on the
candidate list.

\subsection{Gate Decoder and Risk-Controlled Termination}

The plain \ac{BP}-20 decoder serves as the gate and runs first, as shown in Fig.~\ref{fig:intro_fig}. This gated decoder does not require any variable parameters (as it is not adjusted in the sequential ensemble) or complex control flow. Thus, it can be implemented very efficiently, leading to high throughput. 
If the gate syndrome is zero, the corresponding hard decision is returned immediately and no further member is evaluated.
Otherwise, we evaluate the row-boosted members sequentially. A member with zero syndrome contributes its hard decision to the candidate list directly, whereas a member with non-zero syndrome is post-processed by \mbox{\ac{OSD}-1}, so that a decoder whose trajectory did not converge can still contribute a valid codeword candidate. Note that the gate output itself is not post-processed. Hence, the candidate list is empty until the first member has been evaluated. 

Fig.~\ref{fig:complexity} shows the average number of layered \ac{BP} iterations and \mbox{\ac{OSD}-1} activations. Both decrease with the \ac{SNR} and approach the complexity of a single \ac{BP} decoder, since the gate resolves an increasing share of the frames. A detailed complexity analysis is given in Sec.~\ref{subsec:complexity}.

After member \(m\), let \(\hat\cv_m\) denote the best candidate seen so far
and \(\hat\cv_{M-1}\) the full-list decision. The residual risk of stopping
after member \(m\) is
\begin{equation}
 p_m=\Pr\{\hat\cv_m\ne\cv\ \wedge\ \hat\cv_{M-1}=\cv\mid\phiv_m\}.
 \label{eq:residualrisk}
\end{equation}
Note that the second event excludes the frames that the full list does not
recover either and the decision is conditioned on $\phiv_m$ only -- in contrast to a decoding problem, no channel information is present.
\(\phiv_m\) is a feature vector that summarizes the evaluated members using four proxies: the metric \(C(\hat\cv_m)/n\) of the current
winner, its gap \(\bigl(\min_{\tilde\cv\in\Lcal_m\setminus\hat\cv_m}
C(\tilde\cv)-C(\hat\cv_m)\bigr)/n\) to the runner-up, the fraction
\(|\Lcal_m|/(m+1)\) of distinct candidates among the processed members, and the consumed list fraction \(m/M\). We regress \(p_m\) on \(\phiv_m\) using a histogram-based gradient-boosting classifier with \(100\) trees, a learning rate of \(0.08\)\footnote{As a proof of concept; cheaper approaches may be used to
estimate \(p_m\).}~\cite{friedman2001gbm}. The depth is selected automatically, subject to a maximum of \(31\) leaves.
Note that \(\hat p_m\) need not be  a calibrated
probability (but can, if desired), since the thresholds below are fitted empirically and therefore
only require \(\hat p_m\) to be monotone in \(p_m\). 
The decoder stops after member \(m\) if
\begin{equation}
    \hat p_m\leqslant \tau_m ,
    \label{eq:tree-dae}
\end{equation}
where the per-member threshold vector \(\tauv=(\tau_0,\dots,\tau_{M-2})\) is
obtained offline from full-list traces by the error-allocation search of \cite{pillet2026dae}. The search distributes a budget of permitted
premature errors over the members such that a target relative \ac{FER} loss
\(\varepsilon\) is not exceeded. The vector has \(M-1\) entries because the
list is exhausted after the last member, which is equivalent to
\(\tau_{M-1}=\infty\). Note that we reuse the approach of \cite{pillet2026dae} verbatim, as no automorphisms are required for it; only an ensemble is. Both the tree and
\(\tauv\) are fixed after training, and neither the features nor
\eqref{eq:tree-dae}.
Note that the gate and the stopping rule solve the same problem at opposite ends of the decoder: the gate decides whether to start the ensemble at all, and the risk estimate decides when to stop it.
As direct baselines, we apply the same threshold search to the channel metric alone, which recovers plain \ac{DET}, and we evaluate a first-valid mode that omits \ac{OSD} and stops at the first converged member.
\subsection{Two-Core Scheduling with Finite Buffer}
\label{sec:twocore}

Fig.~\ref{fig:architecture} shows the mapping of the gate and the afterburner
onto two cores. Both cores run the same layered \ac{NMSA} kernel. Therefore,
we measure time in \emph{decoder slots}, where one slot corresponds to a
single decoding attempt, i.e., one gate decoding or one ensemble member with $I_\mathrm{max}$ iterations.
Frames arrive periodically, one every \(T_\mathrm{a}\) slots.

Note that \(T_\mathrm{a}\) is a throughput budget rather than a deadline. The
afterburner runs between arrivals, and hence a frame that requires \(d\)
members occupies the second core for \(d\) slots, independently of
\(T_\mathrm{a}\). The first core completes each gate decoding within one slot
and is therefore never congested for \(T_\mathrm{a}\ge1\). The second core
serves gate failures in arrival order.

A buffer of \(B\) frames is placed between the two cores. A gate failure that
finds the afterburner busy waits in this buffer. If the buffer is full on
arrival, the frame currently in afterburner-service is terminated early and emits the best
candidate of its partial list. The head-of-buffer frame then enters service,
and the arriving frame takes the released buffer slot. No frame is dropped,
and hence a finite buffer costs reliability rather than throughput. We report
the \ac{FER} penalty \(\mathrm{FER}(T_\mathrm{a})/\mathrm{FER}_0\), where
\(\mathrm{FER}_0\) is the \ac{FER} of the same decoder with an unbounded buffer.

Let \(\sv_0\) denote the gate syndrome. The mean afterburner service time per
frame is
\begin{equation}
    T_\mathrm{a}^{\ast}
    = \Pr\{\sv_0\neq\zerov\}\;
      \mathbb{E}\!\left[d \mid \sv_0\neq\zerov\right],
    \label{eq:load}
\end{equation}
where \(d\) is the number of afterburner members the stopping rule requests
for a frame, so that \(T_\mathrm{a}^{\ast}\) equals the average number of
processed members minus one (the gate). For
\(T_\mathrm{a}\leqslant T_\mathrm{a}^{\ast}\), the afterburner is overloaded on
average. 

\subsection{Implementation Cost}
\label{sec:cost}
{The objective of risk-controlled termination is to reduce average work by avoiding complete member evaluations, without
changing the constituent decoder or the post-processor. The member-to-member changes for \ac{RBE} are limited to \(\Bcal_m\) and \(\beta_m\) (see Fig.~\ref{fig:architecture}), resulting in low implementation overhead.}
Since \eqref{eq:rowboost} admits both a parallel and a sequential
realization, we report three algorithmic measures separately: the \emph{\ac{BP}
latency}, i.e., the maximum member iteration count for a parallel realization
and the sum for a sequential one; the \ac{BP}-updates computed, i.e., the
completed layered sweeps times the number \(E\) of nonzero entries of
\(\Hm\); and the \ac{OSD} work, i.e., the number of activations, which
we keep separate because the relative cost of the sorter and the \ac{GE} is
architecture-dependent. Implementation overhead from evaluation of the termination criterion is neglected.

\section{Results}
\label{sec:results}

\begin{figure}[!t]
    \centering
    \input{fig/wifi_length_rate_waterfalls.tikz}
    \vspace{-0.6cm}
    \caption{\ac{FER} of four IEEE~802.11 \ac{LDPC} codes under layered
    \acs{NMSA} over \ac{AWGN}. For ensembles, \mbox{\acs{OSD}-1} is applied to
    invalid member outputs. The \ac{DET} decoder matches the performance of the full ensemble.}
    \label{fig:wifi-grid}
    \vspace{-0.2cm}
\end{figure}
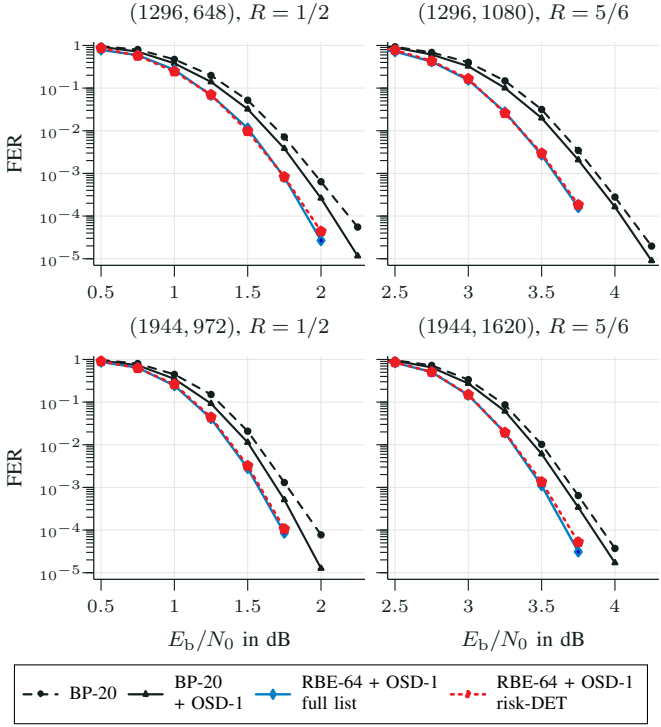

\begin{table*}[t]
    \centering
    \caption{Cost of layered \((1944,1620)\) decoders at
    \(3.5\,\mathrm{dB}\). The parallel ensemble waits for its last member, leading to higher avg. latency.
    }
    \label{tab:cost}
    \resizebox{\textwidth}{!}{\input{fig/wifi1944_layered_cost_table.tex}}
\end{table*}

\subsection{Simulation Setup}
\label{sec:setup}

All schemes share the constituent decoder of Sec.~\ref{sec:decoder}-A with
\(I_{\max}=20\). We fit the normalization factor to \(\alpha=0.825\) at \(R=1/2\) and
\(\alpha=0.80\) at \(R=5/6\). Every ensemble comprises \(M=64\) members, i.e.,
the plain decoder and \(63\) single-row boosts with \(\beta=4\). Its worst
case of \(64\cdot20=1280\) iterations defines the long \ac{BP} baseline.
The optimization \ac{SNR} of a code is the first grid point at which the full list
attains an \ac{FER} of \(10^{-3}\) or below. At this point, we fit one tree and one threshold vector \(\tauv\) per code, using a nominal relative
\ac{FER}-loss target of \(\varepsilon=7\,\%\). Training, calibration, and evaluation use disjoint noise realizations.

\vspace{-0.05cm}
\subsection{Reliability}
Fig.~\ref{fig:wifi-grid} shows the \ac{FER} for different codes. Applying
\mbox{\ac{OSD}-1} to a failed \ac{BP}-20 posterior improves the \ac{FER} by
about an factor of $5$ at rate \(1/2\) and by a factor of $2.5$ at rate \(5/6\).

The ensemble adds a diversity gain on top. On the \((1944,1620)\) code at
\(3.5\,\mathrm{dB}\), the full refined \acs{RBE}-64 improves the \ac{FER} by
almost an order of magnitude over \ac{BP}-20, whereas
\mbox{\ac{BP}-20+\ac{OSD}-1} alone recovers less than a factor of two of that
gain. Risk-\ac{DET} tracks the full list at each design point while
processing at most \(1.15\) members on average, i.e., at most \(15\,\%\) more
\ac{BP} invocations than the gate alone.

The accuracy of the termination estimate varies across the codes. Two of
  the four codes in Fig.~\ref{fig:wifi-grid} stay within \(8\,\%\) of the
  full-list \ac{FER}, whereas the \((1944,972)\) code, for example, exceeds it by
  \(31\,\%\) for target \(\varepsilon=7\,\%\).
Note that the threshold search places every \(\tau_m\)
at the largest value for which the constraint still holds on the training set.
Hence, the validation \ac{FER} loss scatters around \(\varepsilon\) rather than matching it.

In the waterfall region, a relative \ac{FER} excess translates into a small
\ac{SNR} loss, since the \ac{FER} decays steeply with the \ac{SNR}. The
full-list waterfalls in Fig.~\ref{fig:wifi-grid} fall by \(3.0\) to \(5.4\)
decades per \(\mathrm{dB}\), and hence even the largest observed excess corresponds to less than \(0.05\,\mathrm{dB}\). We therefore accept the increase in \ac{FER} compared to the training set. If a tighter bound is required, a larger training set tightens it.

\begin{table*}[t]
    \centering
    \caption{Cost of the layered \((648,540)\) decoders at
    \(3.75\,\mathrm{dB}\).}
    \label{tab:cost-648}
    \resizebox{\textwidth}{!}{\input{fig/wifi648_layered_cost_table.tex}}
    \vspace{-0.1cm}
\end{table*}

\subsection{Average and Worst-Case Cost}
\label{subsec:complexity}
Fig.~\ref{fig:complexity} and Tables~\ref{tab:cost}
and~\ref{tab:cost-648} quantify the cost of the reliability gain. The two
panels of Fig.~\ref{fig:complexity} show the average number of layered \ac{BP} iterations and the number of \mbox{\ac{OSD}-1} activations for \ac{BP}-20, the full refined list, and the two gated approaches. Both gated curves decrease with the
\ac{SNR}, since the gate resolves an increasing share of the frames, whereas
the full list still executes all members (but stops them early on convergence).

The gate dominates the average cost. On \((1944,1620)\) at
\(3.5\,\mathrm{dB}\), it activates the ensemble on only \(1.05\,\%\) of the
frames, and the gated decoder therefore spends about \(46\times\) fewer iterations
than the full list while using a single \ac{BP} instance. A
parallel realization of the full list attains a comparable average latency,
but requires \(64\) \ac{BP} and \(64\) \ac{OSD} datapaths.

The benefit of the partial list features depends on the code. On \((1944,1620)\),
where both rules are fitted at \(3.5\,\mathrm{dB}\) on the same corpus,
risk-\ac{DET} saves \(4\,\%\) of the iterations and \(10\,\%\) of the
\ac{OSD} activations against plain \ac{DET} at the same \ac{FER} (see Tab.~\ref{tab:cost-648}). On
\((648,540)\) at \(3.75\,\mathrm{dB}\), however, it reduces the iteration count by
\(39\times\) and the \ac{OSD} activations by \(18\times\) against the full list, and it improves on plain \ac{DET} by \(20\,\%\) in computed \ac{BP}-updates and by \(30\,\%\) in \ac{OSD} activations at the same \ac{FER} (see Tab.~\ref{tab:cost-648}).

At the matched worst-case budget, \ac{BP}-1280 remains about four times above the full refined list in \ac{FER}, so the gain stems from the diversity of the members and not from a longer trajectory. The first-valid mode is similarly outperformed, since a misconverged  decoder produces a frame error, whereas the full ensemble could still recover the codeword.

The allowed \ac{FER} loss trades reliability for complexity. On a flooding-decoded \((648,540)\) code, accepting an
minor \ac{SNR} loss of less than \(0.01\,\mathrm{dB}\) at \ac{FER} \(10^{-3}\) reduces the average list from \(64\) to \(1.46\) members, whereas the same loss at
\ac{FER} \(10^{-1}\) still requires \(12.20\) members. Note that little remains to be saved where the gate already resolves
most frames: at the optimization points above the average is $1.15$ members, so relaxing the target removes at most a further $15\,\%$.

\subsection{Two-Core Scheduling}
\begin{figure}[t]
    \centering
    \input{fig/two_core_queue.tikz}
    \vspace{-2em}
    \caption{\ac{FER} penalty of the two-core pipeline on the \((648,540)\)
    code under \ac{AWGN}, versus arrival period \(T_\mathrm{a}\) and different buffer depths \(B\). \(\mathrm{FER}_0\) is the same decoder with an
    unbounded buffer. One buffered frame cuts the worst-case penalty from
    \(1.94\) to \(1.14\) at \(3.75\,\mathrm{dB}\)  (\textcolor{mittelblau}{blue}) but
    barely helps at lower \ac{SNR} (\textcolor{orange}{orange}), where the afterburner is overloaded below \(T_\mathrm{a}^{\ast}\).}
    \label{fig:queue-awgn}
\end{figure}
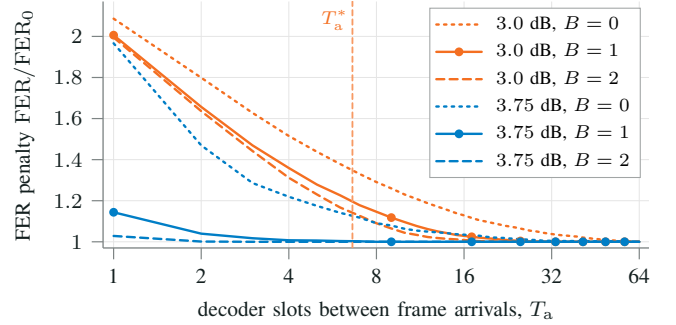
Fig.~\ref{fig:queue-awgn} reports the cost of restricting the decoder to two
cores. At the optimization point, the offered load of \eqref{eq:load} is
\(T_\mathrm{a}^{\ast}=0.14\) slots per frame. A buffer of a single frame
therefore limits the \ac{FER} penalty to \(1.14\) even at \(T_\mathrm{a}=1\),
i.e., at the highest arrival rate that the gate sustains, and the penalty
vanishes from \(T_\mathrm{a}=8\) onwards. At an \ac{FER} of \(10^{-1}\),
however, the load rises to \(T_\mathrm{a}^{\ast}=6.62\), so the afterburner is overloaded on average. In this regime, the penalty at \(T_\mathrm{a}=1\) is
insensitive to the buffer depth \(B\),  since no finite buffer can compensate a mean load that exceeds the service rate. 
Note that the average member count of Sec.~\ref{sec:results}-C therefore determines the admissible arrival rate through \eqref{eq:load}.
Fading relaxes the requirement rather than tightening it: at a matched
\ac{FER} of \(10^{-1}\), the congestion excess over flat Rayleigh block
fading stays two orders of magnitude below the \ac{AWGN}
excess.
\footnote{We will discuss Rayleigh fading in more detail in the final submission, which is not limited to 6 pages.}
Hence, the buffer must be dimensioned for the \ac{AWGN} case.

\section{Conclusion}
\label{sec:conclusion}

This work uses \ac{DET} based on the residual-risk of premature termination for sequential ensembles whose component outputs are all valid
candidates.
Rather than relying on \ac{BP} convergence alone, an \ac{SNR}-blind estimate decides after each member whether an unseen member
could still correct the current decision.
We use the framework with layered \ac{BP}, an \ac{RBE}-64,
conditional \mbox{\ac{OSD}-1}, and a conventional decoder gate on five
IEEE~802.11 \ac{LDPC} codes.
At the optimization points, the rule processes between \(1.04\) and \(1.15\)
members. On the \((648,540)\) code, it reduces \ac{BP} work by \(39\times\)
and \ac{OSD} activations by \(18\times\) relative to the full list.
The load of the afterburner is the average member count minus one.
Hence, the same measurement directly translates to a two-core implementation: A buffer of a single frame limits the congestion penalty to
\(14\,\%\) for every arrival rate that the gate sustains. Note that \ac{AWGN},
rather than fading, is the binding case for buffer design. Fixed-point
evaluation and a channel-matched refit of the risk features remain as future
work.
\vspace{-0.1cm}
\bibliographystyle{IEEEtran}
\bibliography{IEEEabrv,references}

\end{document}

%% file: corporateColours.tex
\definecolor{mittelblau}{RGB}{0, 126, 198}
\definecolor{violettblau}{cmyk}{0.9, 0.6, 0, 0}
\definecolor{rot}{RGB}{238, 28 35}
\definecolor{apfelgruen}{RGB}{140, 198, 62}
\definecolor{gelb}{RGB}{255, 229, 0}
\definecolor{orange}{RGB}{244, 111, 33}
\definecolor{pink}{RGB}{237, 0, 140}
\definecolor{lila}{RGB}{128, 10, 145}
\definecolor{hellgrau}{RGB}{224, 224, 224}
\definecolor{mittelgrau}{RGB}{128, 128, 128}
\definecolor{dunkelgrau}{RGB}{80,80,80}
\definecolor{anthrazit}{RGB}{19, 31, 31}
\definecolor{darkgreen}{RGB}{34,139,34}
\definecolor{aqua}{RGB}{0, 255, 255}

\definecolor{lightgray}{RGB}{211,211,211}

\definecolor{neuesgruen}{RGB}{61, 173, 65}
\definecolor{dunklereshellgrau}{RGB}{176, 176, 176}
\definecolor{neuesgelb}{RGB}{255,160,0}
\definecolor{neuescyan}{RGB}{69,185,224}

\definecolor{tollesgruen}{RGB}{0,217,171}
\definecolor{tollesmagenta}{RGB}{197,67,143}

\definecolor{tollesgelb}{RGB}{255,199,95}
\definecolor{tollesrot}{RGB}{255,111,145}

\colorlet{R12}{apfelgruen}
\colorlet{R23}{mittelblau}
\colorlet{R45}{pink}

%% file: macros.tex
\renewcommand{\vec}[1]{\boldsymbol{#1}}
\newcommand{\vecs}[1]{\boldsymbol{#1}}

\newcommand{\bv}{\vec{b}}
\newcommand{\cv}{\vec{c}}

\newcommand{\sv}{\vec{s}}

\newcommand{\zerov}{\vec{0}}

\newcommand{\Gm}{\vec{G}}
\newcommand{\Hm}{\vec{H}}

%% file: fig/model1.tex
\begin{tikzpicture}[
    xscale=0.85,
    yscale=0.9,
    font=\footnotesize,
    >={Triangle[length=1.3mm, angle'=60]},
]

\draw[line width=1.5pt, dashed, orange, fill=orange!15, rounded corners=3pt] (0.75,-1.25) rectangle (9.75,-6.5);

\node[anchor=south east, orange, align=right] at (9.75,-6.5) {\large \textbf{Afterburner}};

\node[circle, fill=black, draw=black, inner sep = 1.7pt] (channel) at (0,0) {};
\node[anchor=south] at (channel.north) {$\Lch$};

\node[draw=mittelblau, fill=mittelblau!30, rounded corners=3pt, rectangle, align=center, thick, inner sep=4pt] (dec1) at (2,0){LBP \\ decoder 1};

\node[shape=diamond, aspect=2.45, inner sep=2pt, align=center, draw=rot, fill=rot!30, thick] (valid1) at (5,0) {$\sv==0$};

\node[align=center, draw, thick, inner sep=4pt, fill=white] (loadnext) at (5,-2) {load next 
\\ ensemble config};

\node[draw=mittelblau, fill=mittelblau!30, rounded corners=3pt, rectangle, align=center, thick, inner sep=4pt] (dec2) at (2,-3){LBP \\ decoder 2};

\node[shape=diamond, aspect=2.45, inner sep=2pt, align=center, draw=rot, fill=rot!30, thick] (valid2) at (2,-5.25) {$\sv==0$};

\node[shape=diamond, aspect=2.45, inner sep=2pt, align=center, draw=rot, fill=rot!30, thick] (valid3) at (8.5,-2) {stop?};

\node[draw=mittelgrau, fill=white, align=left, thick, rounded corners=3pt, minimum width=4em, minimum height=5em] (list) at (5,-4) {};

\node[draw=orange, fill=orange!30, align=center, thick, rounded corners=3pt, inner sep=4pt] (risk) at (8.5,-4) {calculate \\ risk};

\node[anchor=south] at (list.north) {cw list};

\draw[line width=3pt, mittelgrau!50] (4.5,-3.5) -- (5.5,-3.5);
\draw[line width=3pt, mittelgrau!50] (4.5,-3.75) -- (5.5,-3.75);
\draw[line width=3pt, mittelgrau!50] (4.5,-4.0) -- (5.5,-4.0);
\draw[line width=3pt, mittelgrau!50] (4.5,-4.25) -- (5.5,-4.25);
\draw[line width=3pt, mittelgrau!50] (4.5,-4.5) -- (5.5,-4.5);

\node[draw=apfelgruen, fill=apfelgruen!30, align=center, thick, rounded corners=3pt, inner sep=4pt] (osd) at (3.75,-6) {OSD};

\node[circle, fill=black, draw=black, inner sep = 1.7pt] (chat) at (8.5,0) {};
\node[anchor=south] at (chat.north) {$\hat{\cv}$};

\draw[->, thick] (channel) -- (dec1);
\draw[->, thick] (dec1) -- (valid1);
\draw[->, thick, dashed] (valid1) -- node[pos=0.4, above, sloped] {no} (loadnext);
\draw[->, thick] (loadnext) -| (dec2);
\draw[->, thick] (dec2) -- (valid2);
\draw[->, thick] (valid2) -| node[pos=0.2, above] {yes} (list);
\draw[->, thick] (list) -- (risk);
\draw[->, thick] (risk) -- node[pos=0.45, above, sloped] {$r_i$} (valid3);

\draw[->, thick, dashed] (valid3) -- node[pos=0.45, above] {no} (loadnext);
\draw[->, thick] (valid2) |- node[pos=0.75, above] {no} (osd);
\draw[->, thick] (osd) -| node[pos=0.75, below, sloped] {$\hat{\cv}_i$} (list);

\draw[->, thick] (valid1) -- node[pos=0.5, above] {yes} (chat);
\draw[->, thick] (valid3) -- node[pos=0.55, above, sloped] {yes} (chat);

\draw[->, thick] (channel) |- (dec2);

\end{tikzpicture}

%% file: fig/model2.tex
\begin{tikzpicture}[
    xscale=0.8,  %
    font=\footnotesize,
    >={Triangle[length=1.3mm, angle'=60]},
    vnode/.style={circle, fill=black, draw=black, inner sep=1pt},
    cnode/.style={rectangle, fill=black, draw=black, inner sep=1pt},
    edge/.style={line width=0.3pt}
]
\node[circle, fill=black, draw=black, inner sep=1.7pt] (channel) at (0.35,0) {};
\node[anchor=south] at (channel.north) {$\Lch$};

\begin{scope}[on background layer]
    \node[draw=mittelblau, fill=mittelblau!20, thick, rounded corners=3pt, minimum width=5em, minimum height=3em] (dec1) at (3,0) {};

    \node[shape=diamond, aspect=2.45, inner sep=1pt, align=center, draw=rot, fill=rot!30, thick] (valid1) at (6,0) {$\sv==0$};

    \node[draw=orange, fill=orange!20, thick, rounded corners=3pt, minimum width=9em, minimum height=3.5em] (burner) at (6,-2.5) {};

    \draw[orange, densely dotted, thick] ($(6.0,-2.5) + (0,1.75em)$) -- ($(6.0,-2.5) + (0,-1.75em)$);

    \node[shape=diamond, aspect=1.45, inner sep=1pt, align=center, draw=rot, fill=rot!30, thick] (validx) at ($(6.25,-2.5) +(0.75,0)$) {$\hat{p}_m\leqslant\tau_m$};

    \node[database, database radius=1.56em, database segment height=0.62em, database bottom segment={fill=apfelgruen!20}, database middle segment={fill=apfelgruen!30}, database top segment={fill=apfelgruen!40}] (db1) at (0.35,-1.5) {};

    \node[database, database radius=1.56em, database segment height=0.62em, database bottom segment={fill=apfelgruen!20}, database middle segment={fill=apfelgruen!30}, database top segment={fill=apfelgruen!40}] (db2) at (9,0) {};

    \node[apfelgruen, anchor=south east, align=left, inner sep=0pt] at (db1.north west) {LLR \\ buffer};

    \node[apfelgruen, anchor=north east, align=center, inner sep=0pt] at (db2.south west) {cw \\ buffer};

    \node[database, database radius=1.56em, database segment height=0.62em, database bottom segment={fill=orange!20}, database middle segment={fill=orange!30}, database top segment={fill=orange!40}] (db3) at (9.0,-4.0) {};

    \node[orange, anchor=east, align=center, outer sep=0pt] at (db3.west) {ensemble \\ config};
     
\end{scope}

\node[circle, fill=black, draw=black, inner sep=1.7pt] (cws) at (9.0,1) {};
\node[anchor=south] at (cws.north) {$\hat{\cv}$};

\node[anchor=south, mittelblau] at (dec1.north) {gate decoder};
\node[anchor=south east, orange, outer sep=0pt, inner sep=0pt] at ($(burner.north east) + (0,0.2em) $) {afterburner};

\begin{scope}[shift={(3,0)},scale=1.3]
    \foreach \i in {0,...,4}
        \node[vnode] (v\i) at ({-0.5 + 0.25*\i}, 0.3) {};
    \foreach \j in {0,...,3}
        \node[cnode] (c\j) at ({-0.5 + \j/3}, -0.3) {};

    \begin{pgfonlayer}{edgelayer}
        \foreach \i in {0,...,4}
            \foreach \j in {0,...,3}
                \draw[edge] (v\i) -- (c\j);
    \end{pgfonlayer}
\end{scope}

\begin{scope}[shift={(5.05,-2.5)}, scale=1.3, name prefix=B-,
              cnodehl/.style={rectangle, fill=orange, draw=orange,
                              inner sep=1.6pt, line width=0.4pt}]
    \def\hlcn{1}
    \foreach \i in {0,...,4}
        \node[vnode] (v\i) at ({-0.5 + 0.25*\i}, 0.3) {};
    \foreach \j in {0,...,3}{
        \ifnum\j=\hlcn
            \node[cnodehl] (c\j) at ({-0.5 + \j/3}, -0.3) {};
        \else
            \node[cnode] (c\j) at ({-0.5 + \j/3}, -0.3) {};
        \fi
    }
    \begin{pgfonlayer}{edgelayer}
        \foreach \i in {0,...,4}
            \foreach \j in {0,...,3}
                \draw[edge] (v\i) -- (c\j);
        \foreach \i in {0,...,4}
            \draw[edge, orange, line width=0.6pt] (c\hlcn) -- (v\i);
    \end{pgfonlayer}
\end{scope}

\draw[->, thick] (channel) -- (dec1);
\draw[->, thick] (dec1) -- (valid1);
\draw[->, thick, dashed] (valid1) -- node[pos=0.45, above, sloped] {no} (burner);
\draw[->, thick] (channel) -- (db1);
\draw[->, thick] (db1) |- (burner);
\draw[->, thick, dashed] (valid1) |- node[pos=0.75, above] {push channel LLR} (db1);
\draw[->, thick] (validx) -| node[pos=0.75, above, sloped] {yes} (db2);
\draw[->, thick] (valid1) -- node[pos=0.45, above, sloped] {yes} (db2);
\draw[->, thick, dashed] (validx.south) -| ++(0,-0.45) -- node[midway,above, pos=0.65] {pop channel LLR} ++(-7.9,0) |- (db1.west);
\draw[->, thick] (db2) -- (cws);

\draw[densely dashed, thick, orange] (burner.south east) -- node[pos=0.3, orange, anchor=south west, inner sep=0pt] {$(\mathcal{B}_i,\beta_i,\mu_i)$} (db3);

\end{tikzpicture}

%% file: fig/wifi1944_layered_complexity.tikz
\begin{tikzpicture}[>={Triangle[length=1.3mm, angle'=60]}]
\begin{groupplot}[
    group style={group size=2 by 1, horizontal sep=18mm},
    width=0.4\linewidth, height=0.24\linewidth,
    ymode=log, log origin=infty, grid=major,
    major grid style={draw=hellgrau!85, line width=0.36pt},
    minor grid style={draw=hellgrau!45, line width=0.22pt},
    axis x line*=bottom, axis y line*=left,
    axis line style={draw=anthrazit, line width=0.50pt},
    tick align=outside, tick style={draw=anthrazit, line width=0.45pt},
    ticklabel style={font=\footnotesize, text=anthrazit},
    label style={font=\footnotesize, text=anthrazit},
    xmin=2.5, xmax=3.7,
    xtick={2.5,2.7,2.9,3.1,3.3,3.5,3.7},
    xlabel={\(E_\mathrm{b}/N_0\) in dB},
    legend cell align={left}, legend columns=1,
    legend style={draw=black, fill=white, font=\scriptsize,
        at={(2.4,0.5)}, anchor=west},
]
\nextgroupplot[ymin=1, ymax=2e3, ylabel={average BP iterations}]
\addplot+[color=anthrazit, mark=*, dashed, mark size=1.25pt, line width=0.9pt]
coordinates {(2.5,19.6152344) (2.6,18.8476562) (2.7,18.2773438) (2.8,15.8066406) (2.9,14.1738281) (3,11.9277344) (3.1,9.48632812) (3.2,7.57617188) (3.3,6.06396484) (3.4,4.91357422) (3.5,4.12160773) (3.6,3.58773744) (3.7,3.20022004) (3.8,2.90925) (3.9,2.68152372) (4,2.49835777)};
\addlegendentry{BP-20}

\draw [thick, ->] (axis cs: 3.65,200) -- node[pos=0.2,left, align=right, font=\footnotesize, inner sep=1pt] {complexity \\ reduction} (axis cs: 3.65,5);

\addplot+[color=mittelblau, solid, mark=diamond*, mark size=1.2pt, line width=0.9pt]
coordinates {(2.5,1257.76758) (2.6,1215.33398) (2.7,1182.23828) (2.8,1036.73633) (2.9,942.916016) (3,791.44043) (3.1,665.710286) (3.2,533.546224) (3.3,431.50013) (3.4,347.922991) (3.5,288.743056) (3.6,247.660283) (3.7,217.700578)};
\addlegendentry{\shortstack[l]{RBE-64 + OSD-1,\\full list}}

\addplot+[color=mittelblau, dotted, mark=diamond*, mark size=1.2pt, line width=0.9pt, mark options={solid}]
coordinates {
        (2.5,1236.890625)
        (2.6,1133.930469)
        (2.7,1019.056250)
        (2.8,850.870313)
        (2.9,672.225781)
        (3.0,445.149554)
        (3.1,296.823958)
        (3.2,160.237813)
        (3.3,76.2229818)
        (3.4,37.6467231)
        (3.5,17.0581765)
        (3.6,8.25521238)
  };
\addlegendentry{\shortstack[l]{gated RBE + OSD-1,\\full list}}

\addplot+[color=neuesgruen, mark=pentagon*, solid, mark size=1.25pt, line width=0.9pt]
coordinates {(2.5,628.048828) (2.6,587.871094) (2.7,523.279297) (2.8,387.855469) (2.9,286.46875) (3,181.629883) (3.1,107.283203) (3.2,57.978125) (3.3,29.4042969) (3.4,12.1831267) (3.5,6.43260467) (3.6,4.27977837) (3.7,3.37236986)};
\addlegendentry{\shortstack[l]{gated RBE + OSD-1,\\plain DET}}

\addplot+[color=rot, solid, mark=pentagon*, mark size=1.1pt, line width=1.0pt,mark options={solid}]
coordinates {
      (2.5,822.482143)
      (2.6,717.484375)
      (2.7,602.064062)
      (2.8,457.294531)
      (2.9,327.870313)
      (3.0,188.990327)
      (3.1,107.148698)
      (3.2,54.4379688)
      (3.3,24.6672526)
      (3.4,11.9172309)
      (3.5,6.28608296)
      (3.6,4.15896099)
      (3.7,3.36216826)
  };
\addlegendentry{\shortstack[l]{gated RBE + OSD-1,\\risk-DET}}

\nextgroupplot[ymin=0.7e-2, ymax=1e2, ytick={1e2,1e1,1e0,1e-1,1e-2}, ylabel={average OSD activations}]
\addplot+[color=mittelblau, solid, mark=diamond*, mark size=1.2pt, line width=0.9pt]
coordinates {(2.5,61.4746094) (2.6,57.28125) (2.7,53.6835938) (2.8,42.2734375) (2.9,35.0664062) (3,25.0888672) (3.1,17.2148438) (3.2,11.031901) (3.3,6.26419271) (3.4,3.2948023) (3.5,1.6931081) (3.6,0.892823559) (3.7,0.477707716)};
\addplot+[color=mittelblau, dotted, mark=diamond*, mark size=1.2pt, line width=0.9pt, mark options={solid}]
coordinates {
        (2.5,61.7142857)
        (2.6,56.3453125)
        (2.7,50.4789063)
        (2.8,41.8929688)
        (2.9,32.8296875)
        (3.0,21.4635417)
        (3.1,14.1361979)
        (3.2,7.4662500)
        (3.3,3.41106771)
        (3.4,1.58198785)
        (3.5,0.619818479)
        (3.6,0.220636433)
  };

\addplot+[color=neuesgruen, mark=pentagon*, solid, mark size=1.25pt, line width=0.9pt]
coordinates {(2.5,31.3730469) (2.6,29.3261719) (2.7,26.0605469) (2.8,19.2226562) (2.9,14.1132812) (3,8.82910156) (3.1,5.09244792) (3.2,2.64257812) (3.3,1.22475962) (3.4,0.388162364) (3.5,0.124436599) (3.6,0.0379299588) (3.7,0.00967905031)};
\addplot+[color=rot, solid, mark=pentagon*, mark size=1.1pt, line width=1.0pt,mark options={solid}]
coordinates {
      (2.5,41.109375)
      (2.6,35.8125)
      (2.7,29.9976562)
      (2.8,22.7015625)
      (2.9,16.1859375)
      (3.0,9.19717262)
      (3.1,5.08958333)
      (3.2,2.4603125)
      (3.3,0.987369792)
      (3.4,0.375130208)
      (3.5,0.117948305)
      (3.6,0.032024321)
      (3.7,0.00921182753)
  };

\draw [thick, ->] (axis cs: 3.65,0.5) -- (axis cs: 3.65,0.025);
  
\end{groupplot}
\end{tikzpicture}

%% file: fig/wifi648_r56_fer.tex
\newcommand{\legendDecoderWidth}{1.3cm}
\newcommand{\legendOSDWidth}{1.3cm}

\newcommand{\threecollegend}[2]{%
    \makebox[\legendDecoderWidth][l]{#1}%
    \makebox[\legendOSDWidth][l]{#2}%
}
\begin{tikzpicture}

\begin{axis}[
    width=\linewidth,
    height=0.72\linewidth,
    xmin=1.95,
    xmax=4.55,
    xtick={2,2.5,3,3.5,4,4.5},
    xlabel={\(E_{\mathrm{b}}/N_0\) [dB]},
    ylabel={FER},
    ymode=log,
    ymin=1e-5,
    ymax=1,
    log origin=infty,
    grid=both,
    major grid style={
        draw=hellgrau!85,
        line width=0.36pt
    },
    minor grid style={
        draw=hellgrau!45,
        line width=0.22pt
    },
    axis x line*=bottom,
    axis y line*=left,
    axis line style={
        draw=anthrazit,
        line width=0.5pt
    },
    tick align=outside,
    tick style={
        draw=anthrazit,
        line width=0.4pt
    },
    tick label style={font=\scriptsize},
    label style={font=\small},
    title style={font=\small},
    scaled ticks=false,
    clip=false,
    legend style={
        at={(0.02,0.02)},
        anchor=south west,
        draw=anthrazit,
        fill=white,
        font=\scriptsize\normalfont,
        cells={anchor=west},
        row sep=0pt,
        inner sep=2.5pt,
        column sep=4pt,
    },
    legend cell align=left,
]
\addlegendimage{empty legend}
\addlegendentry{%
    \threecollegend
        {decoder}
        {post-process.}
}
\addplot+[
    color=anthrazit,
    dashed,
    mark=*,
    mark size=1.0pt,
    line width=0.8pt,
    mark options={solid},
] coordinates {
    (2,0.98828125)
    (2.25,0.9453125)
    (2.5,0.87109375)
    (2.75,0.671875)
    (3,0.4361979167)
    (3.25,0.2311197917)
    (3.5,0.08693910256)
    (3.75,0.02473958333)
    (4,0.005715527823)
    (4.25,0.0009929869046)
    (4.5,0.0001410674851)
};
\addlegendentry{%
    \threecollegend{BP}{--}
}

\addplot+[
    color=anthrazit,
    solid,
    mark=triangle*,
    mark size=1.4pt,
    line width=0.8pt,
    mark options={solid},
] coordinates {
    (2,0.970703125)
    (2.25,0.884765625)
    (2.5,0.734375)
    (2.75,0.521484375)
    (3,0.30078125)
    (3.25,0.1438802083)
    (3.5,0.04487179487)
    (3.75,0.01215277778)
    (4,0.002608428805)
    (4.25,0.0004072249326)
    (4.5,5.099741816e-05)
};
\addlegendentry{%
    \threecollegend{BP}{OSD-1}
}

\addplot+[
      color=neuesgruen,
      dashed,
      mark=+,
      mark size=2pt,
      line width=0.9pt,
      mark options={solid},
  ] coordinates {
      (2.0,0.984375)
      (2.5,0.718750)
      (3.0,0.315625)
      (3.5,0.055921053)
      (4.0,0.0026663823)
  };

\addlegendentry{%
    \threecollegend{SMS~\cite{wehn2016afterburner}}{--}
}

\addplot+[
      color=mittelblau,
      dashed,
      mark=diamond*,
      mark size=1.4pt,
      line width=0.9pt,
      mark options={solid},
  ] coordinates {
      (2.00,0.97430556)
      (2.25,0.89513889)
      (2.50,0.73819444)
      (2.75,0.52986111)
      (3.00,0.29444444)
      (3.25,0.13368056)
      (3.50,0.042013889)
      (3.75,0.008984375)
      (4.00,0.0013359788)
  };
\addlegendentry{%
    \threecollegend{RBE~\cite{bezner2026rbe}}{--}
}

\addplot+[
      color=magenta,
      dotted,
      mark=diamond*,
      mark size=1.4pt,
      line width=0.9pt,
      mark options={solid},
  ] coordinates {
      (2.00,0.97430556)
      (2.25,0.89513889)
      (2.50,0.73819444)
      (2.75,0.52986111)
      (3.00,0.29444444)
      (3.25,0.13368056)
      (3.50,0.042013889)
      (3.75,0.008984375)
      (4.00,0.0013359788)
  };
  \addlegendentry{%
    \threecollegend{RBE}{Sys. reenc.}
}

\addplot+[
    color=mittelblau,
    solid,
    mark=diamond*,
    mark size=1.4pt,
    line width=0.9pt,
    mark options={solid},
] coordinates {
    (2,0.859375)
    (2.25,0.671875)
    (2.5,0.49609375)
    (2.75,0.216796875)
    (3,0.09244791667)
    (3.25,0.02514648438)
    (3.5,0.00534119898)
    (3.75,0.0007483237548)
    (4,8.018988966e-05)
};
\addlegendentry{%
    \threecollegend{RBE}{OSD-1}
}

\addplot+[
    color=yellow!50!orange,
    mark=x,
    mark size=2.0pt,
    dotted,
    line width=0.9pt,
    mark options={solid},
] coordinates {
    (2.00,0.94166667)
    (2.25,0.82916667)
    (2.50,0.67708333)
    (2.75,0.39583333)
    (3.00,0.20416667)
    (3.25,0.073263889)
    (3.50,0.019791667)
    (3.75,0.0035663842)
    (4.00,0.0005568484)
};
\addlegendentry{%
    \threecollegend{RBE, DET}{OSD-0}
}

\addplot+[
    color=orange,
    dotted,
    mark=x,
    mark size=2pt,
    line width=1.0pt,
    mark options={solid},
] coordinates {
    (2.00,0.921875)
    (2.25,0.78645833)
    (2.50,0.59270833)
    (2.75,0.28333333)
    (3.00,0.11979167)
    (3.25,0.033333333)
    (3.50,0.0077546296)
    (3.75,0.0012316176)
    (4.00,0.00013616558)
};
\addlegendentry{%
    \threecollegend{RBE, DET}{OSD-1 trunc.}
}

\addplot+[
    color=rot,
    solid,
    mark=x,
    mark size=2pt,
    line width=1.0pt,
    mark options={solid},
] coordinates {
    (2,0.9040178571)
    (2.25,0.7834821429)
    (2.5,0.5982142857)
    (2.75,0.2712053571)
    (3,0.1091269841)
    (3.25,0.0283203125)
    (3.5,0.00623139881)
    (3.75,0.0008892999431)
    (4,0.0001024758157)
};

\addlegendentry{%
    \threecollegend{RBE, DET}{OSD-1}
    }

\end{axis}
\end{tikzpicture}

%% file: fig/wifi_length_rate_waterfalls.tikz
\begin{tikzpicture}
\begin{groupplot}[
    group style={
        group size=2 by 2,
        horizontal sep=3mm,
        vertical sep=12mm,
        xlabels at=edge bottom,
        ylabels at=edge left,
        yticklabels at=edge left,
    },
    width=0.285\textwidth,
    height=0.25\textwidth,
    ymode=log,
    log origin=infty,
    ymin=7e-6,
    ymax=1.2,
    grid=major,
    major grid style={draw=hellgrau!85, line width=0.34pt},
    axis x line*=bottom,
    axis y line*=left,
    axis line style={draw=anthrazit, line width=0.48pt},
    tick align=outside,
    tick style={draw=anthrazit, line width=0.42pt},
    ticklabel style={font=\scriptsize, text=anthrazit},
    label style={font=\footnotesize, text=anthrazit},
    title style={font=\footnotesize, text=anthrazit, yshift=-1mm},
    xlabel={\(E_\mathrm{b}/N_0\) in dB},
    every axis plot/.append style={
        line join=round,
        line cap=round,
    },
    ylabel={FER},
    ytick={1e-5,1e-4,1e-3,1e-2,1e-1,1},
    yticklabels={\(10^{-5}\),\(10^{-4}\),\(10^{-3}\),\(10^{-2}\),\(10^{-1}\),\(1\)},
    yticklabel style={
        font=\tiny,
        text=anthrazit,
        inner sep=0pt,
    },
    ylabel style={
        font=\footnotesize,
        text=anthrazit,
        yshift=3pt,
    },
]

\nextgroupplot[
    title={\((1296,648)\), \(R=1/2\)},
    xmin=0.45,
    xmax=2.3,
    xtick={0.5,1,1.5,2},
    legend to name=wifiFerLegend,
    legend columns=4,
    legend cell align=left,
    legend style={
        draw=black,
        fill=white,
        font=\scriptsize,
        inner xsep=3pt,
        inner ysep=2pt,
        column sep=2pt,
        /tikz/every odd column/.append style={column sep=2pt},
    },
]
\addlegendimage{
    color=anthrazit,
    dashed,
    mark=*,
    mark size=1.1pt,
    line width=0.85pt,
    legend image post style={xscale=0.75},
}
\addlegendentry{BP-20}

\addlegendimage{
    color=anthrazit,
    solid,
    mark=triangle*,
    mark size=1.2pt,
    line width=0.85pt,
    legend image post style={xscale=0.75},
}
\addlegendentry{\shortstack[l]{BP-20\\+ OSD-1}}

\addlegendimage{
    color=mittelblau,
    solid,
    mark=diamond*,
    mark size=1.7pt,
    line width=0.9pt,
    legend image post style={xscale=0.75},
}
\addlegendentry{\shortstack[l]{RBE-64 + OSD-1\\full list}}

\addlegendimage{
    color=rot,
    dotted,
    mark=pentagon*,
    mark size=1.2pt,
    line width=1.0pt,
    legend image post style={xscale=0.75},
}
\addlegendentry{\shortstack[l]{RBE-64 + OSD-1\\risk-DET}}
\addplot+[
    color=anthrazit,
    dashed,
    mark=*,
    mark size=1.0pt,
    line width=0.8pt,
    forget plot,
        mark options={solid}
]
coordinates {
    (0.5,0.9548611111)
    (0.75,0.8020833333)
    (1,0.4713541667)
    (1.25,0.1979166667)
    (1.5,0.05193865741)
    (1.75,0.007148203593)
    (2,0.0006387294574)
    (2.25,5.523497609e-05)
};

\addplot+[
    color=anthrazit,
    solid,
    mark=triangle*,
    mark size=1.1pt,
    line width=0.8pt,
    forget plot,
        mark options={solid}
]
coordinates {
    (0.5,0.9305555556)
    (0.75,0.7152777778)
    (1,0.3802083333)
    (1.25,0.1399739583)
    (1.5,0.03211805556)
    (1.75,0.003836077844)
    (2,0.0002619435957)
    (2.25,1.160398657e-05)
};

\addplot+[
    color=mittelblau,
    solid,
    mark=diamond*,
    mark size=1.7pt,
    line width=0.9pt,
    forget plot,
]
coordinates {
    (0.5,0.7878787879)
    (0.75,0.5916666667)
    (1,0.2684210526)
    (1.25,0.06824324324)
    (1.5,0.01149425287)
    (1.75,0.0008093234056)
    (2,2.66961392e-05)
};

\addplot+[
    color=rot,
    dotted,
    mark=pentagon*,
    mark size=1.7pt,
    line width=1.0pt,
    forget plot,
        mark options={solid}
]
coordinates {
    (0.5,0.8660714286)
    (0.75,0.5736607143)
    (1,0.2455357143)
    (1.25,0.07005494505)
    (1.5,0.009802018634)
    (1.75,0.0008282533793)
    (2,4.386888468e-05)
};

\nextgroupplot[
    title={\((1296,1080)\), \(R=5/6\)},
    xmin=2.45,
    xmax=4.3,
    xtick={2.5,3,3.5,4},
]
\addplot+[
    color=anthrazit,
    dashed,
    mark=*,
    mark size=1.0pt,
    line width=0.8pt,
    forget plot,
        mark options={solid}
]
coordinates {
    (2,1)
    (2.25,0.98828125)
    (2.5,0.9296875)
    (2.75,0.685546875)
    (3,0.40234375)
    (3.25,0.1477272727)
    (3.5,0.03152095376)
    (3.75,0.003450858161)
    (4,0.0002787982157)
    (4.25,1.973125935e-05)
};

\addplot+[
    color=anthrazit,
    solid,
    mark=triangle*,
    mark size=1.1pt,
    line width=0.8pt,
    forget plot,
        mark options={solid}
]
coordinates {
    (2,1)
    (2.25,0.970703125)
    (2.5,0.892578125)
    (2.75,0.611328125)
    (3,0.3255208333)
    (3.25,0.1003787879)
    (3.5,0.0198699422)
    (3.75,0.002074562824)
    (4,0.0001639989504)
    (4.25,8.989184214e-06)
};

\addplot+[
    color=mittelblau,
    solid,
    mark=diamond*,
    mark size=1.7pt,
    line width=0.9pt,
    forget plot,
]
coordinates {
    (2,0.9833333333)
    (2.25,0.9166666667)
    (2.5,0.71875)
    (2.75,0.4166666667)
    (3,0.1548611111)
    (3.25,0.02750544662)
    (3.5,0.002713869086)
    (3.75,0.0001590482552)
};

\addplot+[
    color=rot,
    dotted,
    mark=pentagon*,
    mark size=1.7pt,
    line width=1.0pt,
    forget plot,
        mark options={solid}
]
coordinates {
    (2,0.9910714286)
    (2.25,0.9508928571)
    (2.5,0.7790178571)
    (2.75,0.431547619)
    (3,0.1651785714)
    (3.25,0.02589285714)
    (3.5,0.00296145686)
    (3.75,0.0001829089845)
};

\nextgroupplot[
    title={\((1944,972)\), \(R=1/2\)},
    xmin=0.45,
    xmax=2.3,
    xtick={0.5,1,1.5,2},
]
\addplot+[
    color=anthrazit,
    dashed,
    mark=*,
    mark size=1.0pt,
    line width=0.8pt,
    forget plot,
        mark options={solid}
]
coordinates {
    (0.5,0.9722222222)
    (0.75,0.8012820513)
    (1,0.4466145833)
    (1.25,0.1508928571)
    (1.5,0.02078804348)
    (1.75,0.001302934925)
    (2,7.721260231e-05)
};

\addplot+[
    color=anthrazit,
    solid,
    mark=triangle*,
    mark size=1.1pt,
    line width=0.8pt,
    forget plot,
        mark options={solid}
]
coordinates {
    (0.5,0.9409722222)
    (0.75,0.7307692308)
    (1,0.3502604167)
    (1.25,0.09241071429)
    (1.5,0.01136775362)
    (1.75,0.0005135096468)
    (2,1.263708712e-05)
};

\addplot+[
    color=mittelblau,
    solid,
    mark=diamond*,
    mark size=1.7pt,
    line width=0.9pt,
    forget plot,
]
coordinates {
    (0.5,0.87109375)
    (0.75,0.64375)
    (1,0.244047619)
    (1.25,0.04108255452)
    (1.5,0.002805206463)
    (1.75,8.521226375e-05)
};

\addplot+[
    color=rot,
    dotted,
    mark=pentagon*,
    mark size=1.7pt,
    line width=1.0pt,
    forget plot,
        mark options={solid}
]
coordinates {
    (0.5,0.90625)
    (0.75,0.6272321429)
    (1,0.2622767857)
    (1.25,0.04400510204)
    (1.5,0.003211716341)
    (1.75,0.000106292517)
};

\nextgroupplot[
    title={\((1944,1620)\), \(R=5/6\)},
    xmin=2.45,
    xmax=4.3,
    xtick={2.5,3,3.5,4},
]
\addplot+[
    color=anthrazit,
    dashed,
    mark=*,
    mark size=1.0pt,
    line width=0.8pt,
    forget plot,
        mark options={solid}
]
coordinates {
    (2,1)
    (2.25,0.9921875)
    (2.5,0.955078125)
    (2.75,0.7265625)
    (3,0.3359375)
    (3.25,0.08634868421)
    (3.5,0.01020346004)
    (3.75,0.0006439804772)
    (4,3.710142469e-05)
};

\addplot+[
    color=anthrazit,
    solid,
    mark=triangle*,
    mark size=1.1pt,
    line width=0.8pt,
    forget plot,
        mark options={solid}
]
coordinates {
    (2,1)
    (2.25,0.9921875)
    (2.5,0.935546875)
    (2.75,0.650390625)
    (3,0.2734375)
    (3.25,0.06112938596)
    (3.5,0.006091617934)
    (3.75,0.0003406317787)
    (4,1.709743073e-05)
};

\addplot+[
    color=mittelblau,
    solid,
    mark=diamond*,
    mark size=1.7pt,
    line width=0.9pt,
    forget plot,
]
coordinates {
    (2,0.9958333333)
    (2.25,0.9875)
    (2.5,0.8402777778)
    (2.75,0.5069444444)
    (3,0.1458333333)
    (3.25,0.01912878788)
    (3.5,0.001101919402)
    (3.75,3.094729675e-05)
};

\addplot+[
    color=rot,
    dotted,
    mark=pentagon*,
    mark size=1.7pt,
    line width=1.0pt,
    forget plot,
        mark options={solid}
]
coordinates {
    (2,1)
    (2.25,0.9955357143)
    (2.5,0.859375)
    (2.75,0.5089285714)
    (3,0.1495535714)
    (3.25,0.0193768997)
    (3.5,0.001340626341)
    (3.75,5.208333333e-05)
};

\end{groupplot}

\node[
    anchor=north,
] at
    ($(group c1r2.south)!0.5!(group c2r2.south)+(-0.6cm,-10mm)$)
    {\ref{wifiFerLegend}};

\end{tikzpicture}

%% file: fig/wifi1944_layered_cost_table.tex
\begin{tabular}{@{}lccccccccc@{}}
\toprule
& \multicolumn{2}{c}{parallel instances} & &
  \multicolumn{2}{c}{latency [iters]} &
  \multicolumn{2}{c}{messages [$10^6$]} &
  \multicolumn{2}{c}{OSD calls}\\
\cmidrule(lr){2-3}\cmidrule(lr){5-6}\cmidrule(lr){7-8}\cmidrule(lr){9-10}
decoder & BP & OSD & FER & avg. & max. & avg. & max. & avg. & max.\\
\midrule
BP-20 & 1 & -- & $1.02{\times}10^{-2}$ & 4.12 & 20 &
0.053 & 0.256 & 0.000 & 0 \\
BP-1280 (\(20{\times}64\)) & 1 & -- & $4.36{\times}10^{-3}$ & 10.53 & 1280 &
0.135 & 16.381 & 0.000 & 0 \\
RBE-64, first valid (no OSD) & 1 & -- & $4.74{\times}10^{-3}$ & 11.81 & 1280 &
0.151 & 16.381 & 0.000 & 0 \\
\noalign{\vskip 1pt}
\hdashline
\noalign{\vskip 1pt}
RBE-64 + OSD-1, full parallel & 64 & 64 &
$\mathbf{1.10{\times}10^{-3}}$ &
12.34 & \textbf{20} & 3.689 & 16.381 & 1.673 & 64 \\
RBE-64 + OSD-1, full sequential & 1 & 1 &
$\mathbf{1.10{\times}10^{-3}}$ &
288.22 & 1280 & 3.689 & 16.381 & 1.673 & 64 \\
Gated RBE + OSD-1, plain DET & 1 & 1 & $1.36{\times}10^{-3}$ &
6.56 & 1280 & 0.084 & 16.381 &
0.131 & 64 \\
\rowcolor{mittelblau!20}
Gated RBE + OSD-1, risk-DET
& 1 & 1 & $1.34{\times}10^{-3}$ & \textbf{6.29} & 1280 &
\textbf{0.080} & 16.381 & \textbf{0.118} & 64 \\
\bottomrule
\end{tabular}

%% file: fig/wifi648_layered_cost_table.tex
\begin{tabular}{@{}lccccccccc@{}}
\toprule
& \multicolumn{2}{c}{parallel instances} & &
  \multicolumn{2}{c}{latency [iters]} &
  \multicolumn{2}{c}{messages [$10^6$]} &
  \multicolumn{2}{c}{OSD calls}\\
\cmidrule(lr){2-3}\cmidrule(lr){5-6}\cmidrule(lr){7-8}\cmidrule(lr){9-10}
decoder & BP & OSD & FER & avg. & max. & avg. & max. & avg. & max.\\
\midrule
BP-20 & 1 & -- & $2.47{\times}10^{-2}$ & 3.09 & 20 &
0.015 & 0.095 & 0.000 & 0 \\
\noalign{\vskip 1pt}
\hdashline
\noalign{\vskip 1pt}
Gated RBE + OSD-1, plain DET & 1 & 1 &
$\mathbf{8.50{\times}10^{-4}}$ &
7.62 & 1280 & 0.036 & 6.083 & 0.246 & 64 \\
\rowcolor{mittelblau!20}
Gated RBE + OSD-1, risk-DET
& 1 & 1 & $8.89{\times}10^{-4}$ & \textbf{6.07} & 1280 &
\textbf{0.029} & 6.083 & \textbf{0.171} & 64 \\
\bottomrule
\end{tabular}

%% file: fig/two_core_queue.tikz
\begin{tikzpicture}
\begin{axis}[
    width=\linewidth, height=0.55\linewidth,
    xmode=log, log basis x=2,
    xmin=0.92, xmax=69.12,
    xtick={1,2,4,8,16,32,64}, xticklabels={1,2,4,8,16,32,64},
    ymin=0.97, ymax=2.17,
    ytick={1,1.2,1.4,1.6,1.8,2,2.2},
    yticklabel style={/pgf/number format/fixed,
        /pgf/number format/precision=1},
    grid=both,
    major grid style={draw=hellgrau!85, line width=0.36pt},
    minor grid style={draw=hellgrau!45, line width=0.22pt},
    axis x line*=bottom, axis y line*=left,
    axis line style={draw=anthrazit!55, line width=0.50pt},
    tick align=outside,
    tick style={draw=anthrazit!55, line width=0.45pt},
    ticklabel style={font=\footnotesize, text=anthrazit},
    label style={font=\footnotesize, text=anthrazit},
    xlabel={decoder slots between frame arrivals, \(T_\mathrm{a}\)},
    ylabel={FER penalty \(\mathrm{FER}/\mathrm{FER}_0\)},
    legend columns=1, legend cell align=left,
    legend style={at={(0.985,0.975)}, anchor=north east,
        draw=anthrazit!40, fill=white, font=\scriptsize,
        inner xsep=2pt, inner ysep=1pt, column sep=3pt},
    every axis plot/.append style={line join=round, line cap=round, mark size=1.2pt, line width=0.9pt},
]

\addplot+[orange, dotted, no marks]
coordinates { (1,2.08650519) (2,1.798993394) (3,1.627870399) (4,1.51525637) (5,1.438030827) (6,1.378892734) (7,1.330449827) (8,1.290814722) (9,1.259987417) (10,1.230261088) (11,1.208398868) (12,1.187794904) (13,1.17080843) (14,1.155394778) (15,1.139981126) (16,1.128499528) (17,1.116231519) (18,1.105536332) (19,1.098773199) (20,1.091852784) (21,1.085404215) (22,1.078641082) (23,1.074237182) (24,1.069833281) (25,1.064485687) (26,1.059138094) (27,1.053947782) (28,1.052060396) (29,1.047813778) (30,1.043881724) (31,1.040421516) (32,1.037905002) (33,1.035231205) (34,1.033186537) (35,1.030984586) (36,1.030198176) (37,1.02831079) (38,1.026737968) (39,1.025636993) (40,1.02327776) (41,1.020603964) (42,1.018716578) (43,1.017930167) (44,1.015570934) (45,1.014469959) (46,1.013368984) (47,1.01163888) (48,1.010852469) (49,1.00943693) (50,1.008178673) (51,1.00723498) (52,1.006763133) (53,1.00581944) (54,1.004875747) (55,1.004561183) (56,1.003302925) (57,1.002673797) (58,1.00220195) (59,1.001887386) (60,1.001572822) (61,1.000943693) (62,1.000471846) (63,1) (64,1) };
\addlegendentry{\(3.0\) dB, \(B=0\)}
\addplot+[orange, solid, mark=*, mark repeat=8, mark phase=1,mark options={solid}]
coordinates { (1,2.006134004) (2,1.6569676) (3,1.469644542) (4,1.359232463) (5,1.278389431) (6,1.226486316) (7,1.178357974) (8,1.146272413) (9,1.117804341) (10,1.09594212) (11,1.076753696) (12,1.061811891) (13,1.050330293) (14,1.04215162) (15,1.033501101) (16,1.028782636) (17,1.024378735) (18,1.020603964) (19,1.017301038) (20,1.01525637) (21,1.011953444) (22,1.009751494) (23,1.007706826) (24,1.005976722) (25,1.003932054) (26,1.003460208) (27,1.002359232) (28,1.002044668) (29,1.001730104) (30,1.001258257) (31,1.000471846) (32,1) (33,1) (34,1) (35,1) (36,1) (37,1) (38,1) (39,1) (40,1) (41,1) (42,1) (43,1) (44,1) (45,1) (46,1) (47,1) (48,1) (49,1) (50,1) (51,1) (52,1) (53,1) (54,1) (55,1) (56,1) (57,1) (58,1) (59,1) (60,1) (61,1) (62,1) (63,1) (64,1) };
\addlegendentry{\(3.0\) dB, \(B=1\)}
\addplot+[orange, densely dashed, no marks]
coordinates { (1,1.995124253) (2,1.636206354) (3,1.442277446) (4,1.311733249) (5,1.230261088) (6,1.167190941) (7,1.126769424) (8,1.090279962) (9,1.064800252) (10,1.043881724) (11,1.033501101) (12,1.02327776) (13,1.018244731) (14,1.014312677) (15,1.011953444) (16,1.008650519) (17,1.006134004) (18,1.00361749) (19,1.002673797) (20,1.002359232) (21,1.001100975) (22,1.000943693) (23,1.000471846) (24,1.000471846) (25,1.000471846) (26,1.000314564) (27,1.000157282) (28,1) (29,1) (30,1) (31,1) (32,1) (33,1) (34,1) (35,1) (36,1) (37,1) (38,1) (39,1) (40,1) (41,1) (42,1) (43,1) (44,1) (45,1) (46,1) (47,1) (48,1) (49,1) (50,1) (51,1) (52,1) (53,1) (54,1) (55,1) (56,1) (57,1) (58,1) (59,1) (60,1) (61,1) (62,1) (63,1) (64,1) };
\addlegendentry{\(3.0\) dB, \(B=2\)}
\addplot+[mittelblau, dotted, no marks]
coordinates { (1,1.966542751) (2,1.469640644) (3,1.286245353) (4,1.219330855) (5,1.17472119) (6,1.141263941) (7,1.11771995) (8,1.090458488) (9,1.079306072) (10,1.063197026) (11,1.054522924) (12,1.049566295) (13,1.045848823) (14,1.042131351) (15,1.037174721) (16,1.034696406) (17,1.030978934) (18,1.029739777) (19,1.024783147) (20,1.021065675) (21,1.019826518) (22,1.018587361) (23,1.016109046) (24,1.014869888) (25,1.012391574) (26,1.011152416) (27,1.009913259) (28,1.008674102) (29,1.007434944) (30,1.004956629) (31,1.004956629) (32,1.004956629) (33,1.004956629) (34,1.004956629) (35,1.003717472) (36,1.003717472) (37,1.003717472) (38,1.003717472) (39,1.003717472) (40,1.003717472) (41,1.003717472) (42,1.003717472) (43,1.003717472) (44,1.002478315) (45,1.002478315) (46,1.002478315) (47,1.002478315) (48,1.002478315) (49,1.002478315) (50,1.002478315) (51,1.002478315) (52,1.002478315) (53,1.002478315) (54,1.002478315) (55,1.002478315) (56,1.002478315) (57,1.002478315) (58,1.001239157) (59,1.001239157) (60,1) (61,1) (62,1) (63,1) (64,1) };
\addlegendentry{\(3.75\) dB, \(B=0\)}
\addplot+[mittelblau, solid, mark=*, mark repeat=8, mark phase=1,mark options={solid}]
coordinates { (1,1.143742255) (2,1.039653036) (3,1.017348203) (4,1.007434944) (5,1.006195787) (6,1.003717472) (7,1.002478315) (8,1) (9,1) (10,1) (11,1) (12,1) (13,1) (14,1) (15,1) (16,1) (17,1) (18,1) (19,1) (20,1) (21,1) (22,1) (23,1) (24,1) (25,1) (26,1) (27,1) (28,1) (29,1) (30,1) (31,1) (32,1) (33,1) (34,1) (35,1) (36,1) (37,1) (38,1) (39,1) (40,1) (41,1) (42,1) (43,1) (44,1) (45,1) (46,1) (47,1) (48,1) (49,1) (50,1) (51,1) (52,1) (53,1) (54,1) (55,1) (56,1) (57,1) (58,1) (59,1) (60,1) (61,1) (62,1) (63,1) (64,1) };
\addlegendentry{\(3.75\) dB, \(B=1\)}
\addplot+[mittelblau, densely dashed, no marks]
coordinates { (1,1.02850062) (2,1.001239157) (3,1) (4,1) (5,1) (6,1) (7,1) (8,1) (9,1) (10,1) (11,1) (12,1) (13,1) (14,1) (15,1) (16,1) (17,1) (18,1) (19,1) (20,1) (21,1) (22,1) (23,1) (24,1) (25,1) (26,1) (27,1) (28,1) (29,1) (30,1) (31,1) (32,1) (33,1) (34,1) (35,1) (36,1) (37,1) (38,1) (39,1) (40,1) (41,1) (42,1) (43,1) (44,1) (45,1) (46,1) (47,1) (48,1) (49,1) (50,1) (51,1) (52,1) (53,1) (54,1) (55,1) (56,1) (57,1) (58,1) (59,1) (60,1) (61,1) (62,1) (63,1) (64,1) };
\addlegendentry{\(3.75\) dB, \(B=2\)}
\addplot[orange!70, dash pattern=on 2pt off 1.5pt, line width=0.7pt, no marks, forget plot]
coordinates { (6.617,0.97) (6.617,2.17) };
\node[anchor=north east, font=\scriptsize, text=orange, inner sep=1.5pt] at (axis cs:6.617,2.15) {\(T_\mathrm{a}^\ast\)};
\end{axis}
\end{tikzpicture}